\documentclass[tightenlines,superscriptaddress,eqsecnum,floats,nofootinbib,showpacs]
{revtex4}
\usepackage{amssymb}
\usepackage{stmaryrd}
\usepackage{amsmath}
\usepackage{amsfonts}
\usepackage{mathrsfs}

\usepackage{amsmath,amssymb,amsfonts}
\usepackage{graphicx}

\renewcommand{\v}{\nu}
\renewcommand{\l}{\lambda}
\newcommand{\f}{\phi}
\newcommand{\p}{p_{\f}}
\newcommand{\G}[4]{G^{#1,#2}_{#3,#4}}

\begin{document}

%\date\today
%\preprint{????}

\title{Effective Theories of Quantum Cosmology}
\author{Xinquan Wu\footnote{wuxq@mail.bnu.edu.cn} and Yongge Ma\footnote{mayg@bnu.edu.cn}}
\affiliation{Department of Physics, Beijing Normal University, Beijing 100875, China}

\begin{abstract}
We introduce two possible ways of defining effective constraints of quantum systems and applied this effective constraint method to models of WDW Quantum Cosmology and Loop Quantum Cosmology. We analyze effective Hamiltonian constraint on both second and third order and calculate Hubble parameter as well as modified Friedmann equation of each model. Then we compare with a special case using coherent state. It shows that this method is reasonable and as before the classical Big Bang singularity is replaced by a quantum bounce in Loop Quantum Cosmology.

\end{abstract}


\maketitle

\section{Introduction}\label{sec:introduction}

In quantum systems, there are many applications which are placed in regions close to classical behavior, where quantum properties must be taken into acount.
So it is often helpful to work with equations of classical type, i.e. sets of
ordinary differential equations for mechanical systems, which are amended by correction
terms resulting from quantum theory. Then from a mathematical point of view, the question
becomes how well the behavior of a wave function subject to a partial differential equation
can be approximated by finitely many variables subject to a system of coupled but ordinary
differential equations.

One useful example is the effective equations of motion of quantum mechanics. Quantum mechanics can be formulated into an intrinsically geometric fashion[\onlinecite{AsSc97}].
Based on this formulation, we can develope a method of effective equations of motion which is a trustworthy tool to avoid some of the mathematical and conceptual
difficulties of quantum theories. Quantum corrections to classical equations of motion are
usually easier to analyze than the behavior of outright quantum states, and they can often
be derived in a manageable way[\onlinecite{BoSk06}]. Effective equations are not only quantum
corrected classical equations, but also provide direct solutions for quantum properties such
as expectations values or fluctuations. While semiclassical regimes play important roles in
providing useful approximation schemes, effective equations present a much more general method.

One of the key issues in quantizations of fundamental theories is the determination
of physical observables. They must satisfy the constraint equations and be invariant under
gauge transformations. For canonical quantum theories, solving constraints is traditionally
done at the state level: one constructs a physical Hilbert space of states annihilated by the
constraint operators and equipped with an invariant inner product. Explicit constructions
can be done in some special cases by different methods.
Since explicit derivations are possible only in specific cases, it is not always clear whether
the results are generic or mere artefacts of the simple models used. It is therefore important
to have approximate methods for a wider range of cases, or at least to be able to
perturb around known solvable ones while still ensuring that the constraints are solved
and the observables are gauge invariant. It turns out that such perturbation schemes are
most feasible if one deals with the observables directly, such as expectation values. This procedure gives rise
to canonical effective equations and constraints[\onlinecite{BoSk06,BoSa09}].

So for constrained systems, we can develop an effective constraint
formalism parallel to that of unconstrained systems. Its advantages
are that (i) it avoids directly writing an integral (or other) form of a physical inner product,
which is instead implemented by reality conditions for the physical variables; and (ii) it directly provides
physical quantities such as expectation values and fluctuations as relational functions of
internal time, rather than computing a whole wave function first and then performing
integrations. These advantages avoid conceptual problems and some technical difficulties
in solving quantum equations. They can also bring out general properties more clearly,
especially in quantum cosmology[\onlinecite{Bo08-3,Bo08-2}].

One of the most fundamental issues in modern physics is to establish the theory of quantum gravity.
Among various approaches, the nonperturbative quantization is a very important candidate.
Based on the ADM Hamiltonian formalism of general relativity(GR)
and the Dirac's generalized Hamiltonian quantization method,
Wheeler and DeWitt proposed a wave functional description of gravitational field. In this quantum geometrodynamical
approach, the whole universe is described by a wave function which is defined on superspace, the space of all 3-metrics
and matter field configurations. The dynamics is encoded in the second-order differential Wheeler-DeWitt(WDW)
equation.

There is an alternative background independent approach which is the loop quantum
gravity(LQG)[\onlinecite{Ro04}-\onlinecite{HaMa07}]. The starting point of LQG is the Hamiltonian connection dynamics
of GR rather than the ADM
formalism. In this framework, GR looks like a gauge field theory with SU(2) as its internal gauge group. By taking the
holonomy of su(2)-connection $A^i_a$ and flux of densitized triad $E^b_j$ as basic variables, the quantum kinematical framework
of LQG has been rigorously constructed, and the Hamiltonian constraint operator can also be well defined to represent
quantum dynamics. Moreover, a few physically significant results, especially the resolution of big bang singularity,
have been obtained in the minisuperspace models of loop quantum cosmology(LQC)[\onlinecite{AsBo03}-\onlinecite{AsSi11}].
The quantum bounce
replacement of big bang and its properties are being studied from different prescriptions of LQC[\onlinecite{YaDi11}].
Effective
equations were also derived in isotropic models[\onlinecite{Ta08},\onlinecite{DiMa08}], which predict evolution
of universe with quantum corrections
and shed new lights on the singularity resolution.

In both WDW quantum cosmology and LQC, by coupling with a massless scalar field, the Hamiltonian constraint
equations can be reformulated as Klein-Gordon-like equations, where the corresponding gravitational Hamiltonian
operators, as multiplications of several self-adjoint operators, are non-symmetric in the kinematical Hilbert space.
The above models of quantum
cosmology provide a good arena to launch the desired investigation. Hence we will study the effecive constraints
in spatially flat isotropic FRW cosmology coupled with a massless scalar field in both WDW and loop quantum cosmology.

We give a brief introduction of effective equations of motion for unconstrained system in section \ref{sec:2} and two definitions of effective constraints in section \ref{sec:3}. Then in the major part of this paper, section \ref{sec:4}, we analyze the second and third order effective constraints in WDW Quantum Cosmology and Loop Quantum Cosmology. Finally, we compare the result with the one in a special case using coherent state in the discussion section.

\section{General Formalism}\label{sec:2}

We first review the setup of effective equations for unconstrained Hamiltonian systems[\onlinecite{BoSk06,BoSk07}],
which we will generalize to systems with constraints in the following section.

We describe a state by its moments rather than a wave function. 
This has the immediate advantage that the description deals directly with quantities of physical interest.
In terms of expectation values, fluctuations and all higher moments,
this structure takes the form of an infinite dimensional phase space whose Poisson relations
are derived from the basic commutation algebra. Dynamics is determined by a Hamiltonian
on this phase space. As a function of all the phase space variables it is obtained by taking
the expectation value of the Hamiltonian operator in a general state and expressing the
state dependence as a dependence on all the moments. Thus, the Hamiltonian operator
determines a function on the infinite dimensional phase space which generates Hamiltonian
evolution.

For an ordinary quantum mechanical system with canonical basic operators
$\hat{q}$ and $\hat{p}$ satisfying $[\hat{q}, \hat{p}] = i\hbar$, we have a phase space
coordinatized by the expectation values $q := \langle \hat{q} \rangle$ and
$p := \langle \hat{p} \rangle$ as well as infinitely many quantum variables
\begin{equation}
\label{equ:QuantumVariable}
G^{a,b} := \langle (\hat{p}-\langle\hat{p}\rangle)^a (\hat{q}-\langle\hat{q}\rangle)^b \rangle_{\textrm{Weyl}},
\end{equation}
for integer $a$ and $b$ such that $a + b \ge 2$, where the totally symmetric ordering is used.
For $a + b = 2$, for instance, this provides fluctuations $(\Delta q)^2 = G^{0,2} = G^{qq}$ and
$(\Delta p)^2 = G^{2,0} = G^{pp}$ as well as the covariance $G^{1,1} = G^{qp}$. As indicated, for
moments of lower orders it is often helpful to list the variables appearing as operators directly.
The symplectic structure is determined through Poisson brackets which follow by the basic rule 
$\{A,B\} = -i\hbar^{-1}\langle[\hat{A},\hat{B}]\rangle$ for any two operators $\hat{A}$ and $\hat{B}$
which define phase space functions $A := \langle\hat{A}\rangle$ and $B := \langle\hat{B}\rangle$.
Moreover, for products of expectation values in the quantum variables one
simply uses the Leibniz rule to reduce all brackets to the elementary ones. General Poisson
brackets between the quantum variables then satisfy the formula
\begin{eqnarray}
\{G^{a,b},G^{c,d}\}
  & = & - \sum_{r,s=0}^{\infty}(-\frac14\hbar^2)^{r+s}
          \sum_{j,k}\binom{a}{j}\binom{b}{k}\binom{c}{k}\binom{d}{j}
          G^{a+c-j-k,b+d-j-k}(\delta_{j,2r+1}\delta_{k,2s}-\delta_{j,2r}\delta_{k,2s+1})
\nonumber\\
  &   & + adG^{a-1,b}G^{c,d-1}-bcG^{a,b-1}G^{c-1,d},
\end{eqnarray}
where the summation of $j$ and $k$ is over the ranges $0 \le j \le \min(a, d)$ and
$0 \le k \le \min(b, c)$, respectively. For low order moments, it is easier to use direct calculations of
Poisson brackets via expectation values of commutators. For second order moments, it is easy to get
\begin{equation}
\{G^{0,2},G^{1,1}\}=2G^{0,2},\quad
\{G^{0,2},G^{2,0}\}=4G^{1,1},\quad
\{G^{1,1},G^{2,0}\}=2G^{2,0}.
\end{equation}
This defines the kinematics of
the quantum system formulated in terms of moments.

Dynamics is defined by a quantum Hamiltonian derived from the Hamiltonian operator
by taking expectation values. This results in a function of expectation values and moments
through the state used for the expectation value. By Taylor expansion, we have
\begin{eqnarray}
\label{equ:hamiltonian}
H_Q(q,p,G^{a,b})
  & = & \langle H(\hat{q},\hat{p})_{\textrm{Weyl}}\rangle
    =   \langle H(q+(\hat{q}-q),p+(\hat{p}-p))_{\textrm{Weyl}}\rangle
\nonumber\\
  & = & H(q,p)+\sum_{a=0}^{\infty}\sum_{b=0}^{\infty}\frac1{a!b!}
        \frac{\partial^{a+b}H(q,p)}{\partial p^a\partial q^b}G^{a,b},
\end{eqnarray}
where we understand $G^{a,b} = 0$ if $a + b < 2$ and $H(q, p)$ is the classical Hamiltonian
evaluated in expectation values. As written explicitly, we assume the Hamiltonian to be
Weyl ordered. If another ordering is desired, it can be reduced to Weyl ordering by adding
re-ordering terms.

Having a Hamiltonian and Poisson relations of all the quantum variables, one can compute
Hamiltonian equations of motion $\dot{q} = \{q,H_Q\}$, $\dot{p} = \{p,H_Q\}$ and $\dot{G}^{a,b} = \{G^{a,b},H_Q\}$.
This results in infinitely many equations of motion which, in general, are all coupled to
each other. This set of infinitely many ordinary differential equations is fully equivalent
to the partial differential equation for a wave function given by the Schr\"odinger equation.
In general, one can expect a partial differential equation to be solved more easily than infinitely
many coupled ordinary ones. Exceptions are solvable systems such as the harmonic
oscillator or the spatially flat quantum cosmology of a free, massless scalar field[\onlinecite{Bo07-1}] where
equations of motion for expectation values and higher moments decouple. In quantum cosmology,
this is developed in [\onlinecite{Bo07-3,Bo08-1,Bo08-2}]. Moreover, semiclassical
and some other regimes allow one to decouple and truncate the equations consistently,
resulting in a finite set of ordinary differential equations. This is easier to solve and
can be exploited to avoid conceptual problems especially
in the context of constrained systems.

\section{Effective Constraints}\label{sec:3}

For a constrained system, the definition of phase space variables (\ref{equ:QuantumVariable}) in addition to expectation
values of basic operators is the same. For several basic variables, copies of independent
moments as well as cross-correlations between different canonical pairs need to be taken
into account. A useful notation, especially for two canonical pairs $(q, p; q_1, p_1)$ as we will
use it later, is
\begin{equation}
G^{a,b}_{c,d}
  \equiv\langle(\hat{p}-p)^a(\hat{q}-q)^b
               (\hat{p_1}-p_1)^c(\hat{q_1}-q_1)^d\rangle_{\textrm{Weyl}}.
\end{equation}
Also here we will, for the sake of clarity, sometimes use a direct listing of operators, as in
$G^{qq} = G^{2,0}_{0,0} = (\Delta q)^2$ or the covariance $G^{q}_{p_1} = G^{0,1}_{1,0}$, for low order moments.

We assume that we have a single constraint $\hat{C}$ in the quantum system and no true
Hamiltonian; cases of several constraints or constrained systems with a true Hamiltonian
can be analyzed analogously. We clearly must impose the principal quantum constraint
$C_Q(q, p,G^{a,b}) := \langle\hat{C}\rangle = 0$ since any physical state $|\Psi\rangle$, whose expectation values and
moments we are computing, must be annihilated by our constraint, $\hat{C}|\Psi\rangle = 0$. Just as
the quantum Hamiltonian $H_Q$ before, the quantum constraint can be written as a function
of expectation values and quantum variables by Taylor expansion as in (\ref{equ:hamiltonian}). However,
this one condition for the phase space variables is much weaker than imposing a Dirac
constraint on states, $\hat{C}|\Psi\rangle = 0$. In fact, a simple counting of degrees of freedom shows that
additional constraints must be imposed: One classical constraint such as $C = 0$ removes
a pair of canonical variables by restricting to the constraint surface and factoring out the
flow generated by the constraint. For a quantum system, on the other hand, we need to
eliminate infinitely many variables such as a canonical pair $(q, p)$ together with all the
quantum variables it defines. Imposing only $C_Q = 0$ would remove a canonical pair but
leave all its quantum variables unrestricted. These additional variables are to be removed
by infinitely many further constraints.

Here we introduce two ways of defining quantum constraints. The first way is to define constraints 
with operators of non-symmetric ordering. There are obvious candidates for these constraints.
If $\hat{C}|\Psi\rangle = 0$ for any physical state,
we can define infinitely many quantum constraints
\begin{eqnarray}
C^{(n)}&:=&\langle\hat{C}^n\rangle=0,\\
C^{(n)}_{f(q,p)}&:=&\langle f(\hat{q},\hat{p})\hat{C}^n\rangle=0,
\label{equ:constraints}
\end{eqnarray}
for positive integer n and arbitrary phase space functions $f(q, p)$. All these expectation
values vanish for physical states, and in general differ from each other on the quantum
phase space. For arbitrary $f(q, p)$, there is an uncountable number of constraints which
should be restricted suitably such that a closed system of constraints results which provides
a complete reduction of the quantum phase space. The form of functions $f(q, p)$ to be
included in the quantum constraint system depends on the form of the classical constraint
and its basic algebra.

We thus have indeed infinitely many constraints, which constitute the basis for our
effective constraints framework. This is to be solved as a classical constrained system,
but as an infinite one on an infinite dimensional phase space. An effective treatment then
requires approximations whose explicit form depends on the specific constraints. At this
point, some caution is required: approximations typically entail disregarding quantum
variables beyond a certain order to make the system finite. Doing so for an order of
moments larger than two results in a Poisson structure which is not symplectic because only
the expectation values form a symplectic submanifold of the full quantum phase space, but
no set of moments to a certain order does. We are then dealing with constrained systems
on Poisson manifolds such that the usual countings of degrees of freedom do not apply.
For instance, it is not guaranteed that each constraint generates an independent flow even
if it weakly commutes with all other constraints which would usually make it first class.
Properties of constrained systems in the more general setting of Poisson manifolds which
need not be symplectic are discussed, e.g., in [\onlinecite{BoSt03}].

Some of quantum constraints
in (\ref{equ:constraints}) are defined as expectation values of non-symmetric operators, thus implying
complex valued constraint functions. We specifically do not order symmetrically in (\ref{equ:constraints})
because this would give rise to terms where some $\hat{q}$ or $\hat{p}$ appear to the right while others
remain to the left. This would not vanish for physical states and therefore not correspond
to a constraint. This may appear problematic, but one should note that this reality
statement is dependent on the kinematical inner product used before the constraints
are imposed. This inner product in general differs from the physical one if zero is in the
continuous part of the spectrum of the constraint and thus reality in the kinematical inner
product is not physically relevant. Moreover, in gravitational theories it is common or
even required to work with constraint operators which are not self-adjoint[\onlinecite{Ko79}], and thus
complex valued constraints have to be expected in general. For physical statements, which
are derived after the constraints have been implemented, only the final reality conditions
of the physical inner product are relevant.

This physical reality can be implemented effectively:
We solve the constraints on the quantum phase space, and then impose the condition
that the reduced quantum phase space be real. We will see explicitly that complex-valued
quantum variables on the unconstrained phase space are helpful to ensure consistency. In
parallel to Hilbert space notation, we call quantum variables (\ref{equ:QuantumVariable}) on the original quantum
phase space kinematical quantum variables, and those on the reduced quantum phase space
physical quantum variables. Kinematical quantum variables are allowed to take complex
values because their reality would only refer to the inner product used on the kinematical
Hilbert space. For physical quantum variables in the physical Hilbert space as usually
defined, on the other hand, reality conditions must be imposed.

It may seem obvious how to avoid the question of reality of the constraints altogether
by using quantum constraints defined as $G^{C^nf(q,p)} = \langle\hat{C}^n\widehat{f(p,q)}\rangle_{\textrm{Weyl}}$
such as $G^{C^nq}$ and $G^{C^np}$
with the symmetric ordering used as in (\ref{equ:QuantumVariable}). This is second way of defining
the effective constraints which are symmetric. Here, the symmetric ordering contained in
the definition of quantum variables must leave $\hat{C}$ intact as a possibly composite operator,
i.e. we have for instance $G^{C,p} = \frac12 \langle \hat{C}\hat{p} + \hat{p}\hat{C} \rangle - Cp$ independently of the functional form of
$\hat{C}$ in terms of $\hat{q}$ and $\hat{p}$. Otherwise it would not be guaranteed that the expectation value
vanishes on physical states. We could not include variables with higher powers of $q$ and
$p$, such as $G^{C^npp}$ as constraints because there would be terms in the totally symmetric
ordering (such as $\hat{p}\hat{C}^n\hat{p}$) not annihilating a physical state. But, e.g., $G^{Cp^2}$ understood as
$\frac12 \langle \hat{C}\hat{p}^2 + \hat{p}^2\hat{C} \rangle - Cp^2$ would be allowed.
The use of such symmetrically ordered variables would imply real quantum constraints.

However, this definition of quantum constraints may seem problematic: The constraints would not form a closed set
and not even be first class. We have, for instance,
\begin{eqnarray}
\{G^{C^nf(q,p)},G^{C^mg(q,p)}\}
  & = & \frac1{4i\hbar}
        \langle[\hat{C}^n\hat{f}+\hat{f}\hat{C}^n,\hat{C}^m\hat{g}+\hat{g}\hat{C}^m]\rangle
\nonumber\\
  &   & - \frac{g}{2i\hbar}\langle[\hat{C}^n\hat{f}+\hat{f}\hat{C}^n,\hat{C}^m]\rangle
        - \frac{C^m}{2i\hbar}\langle[\hat{C}^n\hat{f}+\hat{f}\hat{C}^n,\hat{g}]\rangle
\nonumber\\
  &   & - \frac{f}{2i\hbar}\langle[\hat{C}^n,\hat{C}^m\hat{g}+\hat{g}\hat{C}^m]\rangle
        - \frac{C^n}{2i\hbar}\langle[\hat{f},\hat{C}^m\hat{g}+\hat{g}\hat{C}^m]\rangle
        + \{C^nf,C^mg\}.
\nonumber
\end{eqnarray}
The first commutator contains several terms which vanish when the expectation value is
taken in a physical state, but also the two contributions $[\hat{C}^n,\hat{g}]\hat{C}^m\hat{f}$ and
$\hat{f}\hat{C}^m[\hat{C}^n,\hat{g}]$ whose
expectation value in a physical state vanishes only if $\hat{f}$ or $\hat{g}$ commute with $\hat{C}$. This would
require quantum observables to be known and used in the quantum constraints, which in
general would be too restrictive and difficult. In some models, we may lucky to find out enough quantum observables.

By contrast, the quantum constraints defined in the first way do form a first class system: We have
\begin{equation}
\label{equ:closure}
[\hat{f}\hat{C}^n,\hat{g}\hat{C}^m]
  = [\hat{f},\hat{g}]\hat{C}^{n+m}
    + \hat{f}[\hat{C}^n,\hat{g}]\hat{C}^m + \hat{g}[\hat{f},\hat{C}^m]\hat{C}^n,
\end{equation}
whose expectation value in any physical state vanishes. Thus, using these constraints
implies that their quantum Poisson brackets vanish on the constraint surface, providing a
weakly commuting set:
\begin{equation}
\{C_f^{(n)},C_g^{(m)}\}
  = \frac1{i\hbar}\langle[\hat{f}\hat{C}^n,\hat{g}\hat{C}^m]\rangle\approx0.
\end{equation}

Constraints thus result for all phase space functions $f(q, p)$, but not all constraints
in this uncountable set can be independent. For practical purposes, one would like to
keep the number of allowed functions to a minimum while keeping the system complete.
Then, however, the set of quantum constraints is not guaranteed to be closed for any
restricted choice of phase space functions in their definition. If $C^{(n)}_f$ and $C^{(m)}_g$ are quantum
constraints, closure requires the presence of $C^{(n)}_{[f,g]}$ (for $n \ge 2$),
$C^{(n)}_{f[C^m,g]}$ and $C^{(n)}_{g[C^m,f]}$ as
additional constraints according to (\ref{equ:closure}). This allows the specification of a construction
procedure for a closed set of quantum constraints. As we will see in examples later, for a
system in canonical variables $(q, p)$ it is necessary to include at least $C^{(n)}_q$ and $C^{(m)}_p$ in the
set of constraints for a complete reduction. With $C^{(n)}_{[q,p]} = i\hbar C^{(n)}$, the first new constraints
resulting from a closed constraint algebra add nothing new. However, in general the new
constraints $C^{(n)}_{q[C^m,p]}$ and $C^{(n)}_{p[C^m,q]}$ will be independent and have to be included. Iteration of
the procedure generates further constraints in a process which may or may not stop after
finitely many steps depending on the form of the classical constraint.

Although many independent constraints have to be considered for a complete system,
most of them will involve quantum variables of a high degree. To a given order in the
moments it is thus sufficient to consider only a finite number of constraints which can
be determined and analyzed systematically.

\section{Effective Constraints of Quantum Cosmology}\label{sec:4}

The Hilbert-Einstein action for these models are given by
\begin{equation}
\label{equ:action}
S = \frac1{16\pi G}\int d^4x\sqrt{-g}R
    + \frac12\int d^4x\sqrt{-g}g^{\mu\nu}\phi_{,\mu}\phi_{,\nu} .
\end{equation}
In this spatially flat model, we fix a space-like sub-manifold $S$, which is
topologically $\mathbb{R}^3$ and equipped with Cartesian 
coordinates $x^i(i = 1, 2, 3)$, and a fiducial flat metric $^oq_{ab}$ given
by $^oq_{ab}dx^adx^b = dx^2_1+dx^2_2+dx^2_3$. The physical 3-metric
$q_{ab}$ is then determined by a scale factor a satisfying $q_{ab} = a^2~^oq_{ab}$.
It is convenient to introduce an elementary cell $\mathcal{V}$ and restrict all integrations to
this cell since the spatial slice is non-compact. The volume of $\mathcal{V}$ with respect to $^oq_{ab}$
is denoted as $V_o$ and the physical volume is $V = a^3V_o$. Then the geometrical
pair $(a, p_a)$ can be used as canonical
variables, where the conjugate momentum satisfies $p_a = - \frac{3V_o}{4\pi G}a\dot{a}$.

In order to study the WDW quantum cosmology and LQC on the same footing, we employ the new canonical variables
$(A^i_a,E^a_i)$ in both theories. Due to the homogeneity and isotropy, we can fix a set of orthonormal cotriad and triad
$(^o\omega^i_a,^o e^a_i)$ compatible with $^oq_{ab}$ and adapted to $\mathcal{V}$. Then the cotriad $\omega^i_a$
which are orthonormal with respect to physical
metric $q_{ab}$ can be written as $\omega^i_a = \chi a^o\omega^i_a$, where $\chi = 1$ if $\omega^i_a$
has the same orientation as the fiducial $^o\omega^i_a$ and $\chi = -1$
if the orientation is opposite. The basic canonical variables take the simple form
\begin{equation}
A_a^i=cV_o^{-\frac13}~^o\omega_a^i,\quad
E_i^a=p\sqrt{^oq}V_o^{-\frac23}~^oe_i^a .
\end{equation}
The dynamical variables are thus reduced to $(c, p)$ with the Poisson bracket: $\{c, p\} = 8\pi G\gamma/3$,
where $\gamma$ is the Barbero-
Immirzi parameter. Following the $\bar{\mu}$-scheme of ``improved dynamics''[\onlinecite{AsPa06-2}],
the regulator $\bar{\mu}$ used in holonomies is given
by $\bar{\mu} = \sqrt{\Delta/|p|}$, where $\Delta = 2\sqrt3\pi\gamma\ell^2_P$ is a minimum nonzero eigenvalue of the area operator. In order to do the
semiclassical analysis, it is convenient to introduce new dimensionless conjugate variables:
\begin{equation}
b:=\frac{c}{2|p|^{\frac12}}=-\frac{2\pi G\gamma}{3V_o}\frac{p_a}{a^2} ,\quad
\nu:=\frac{\textrm{sgn}(p)|p|^{\frac32}}{2\pi G\gamma}=\frac{\epsilon V_oa^3}{2\pi G\gamma} .
\end{equation}
with the Poisson bracket ${b, \v} = 1$, where the Planck length $\ell_P$ is given by $\ell_P^2 = G\hbar$.
From the matter part of
action (\ref{equ:action}), we can get the momentum of $\phi$ as $p_{\phi} = p^{\frac32}\dot{\phi} = a^3V_o\dot{\phi}$
and the Poisson bracket: $\{\phi, p_{\phi}\} = 1$.

\subsection{Effective Constraints of WDW Quantum Cosmology}

Now we apply the effective constraint method to WDW quantum cosmology, considering $k=0$ FRW cosmology model.
We start from the classical Hamiltonian constraint which can be obtained from the action (\ref{equ:action}),
\begin{equation}
C=-\frac{3V_o}{8\pi G}a\dot{a}^2+\frac1{2V_o}\frac{p_{\phi}^2}{a^3}
 =-\frac{3}{8\pi G\gamma^2}c^2|p|^{\frac12}+\frac12\frac{p_{\phi}^2}{|p|^{\frac32}}
 =-\frac3{\gamma}b^2\v+\frac1{4\pi G\gamma}\frac{p_{\phi}^2}{\v}.
\end{equation}
After quantization, we have the corresponding constraint operator
\begin{equation}
\label{equ:constr_oper}
\hat{C}=-\frac3{\gamma}\hat{b}\hat{\v}\hat{b}
        +\frac1{4\pi G\gamma}\frac{\hat{p}_{\phi}^2}{\hat{\v}}
       =A\hat{p}_{\phi}^2\hat{\v}^{-1}
        -B\hat{b}\hat{\v}\hat{b},
\end{equation}
where $A=\frac1{4\pi G\gamma}$, $B=\frac3{\gamma}$, and the operator ordering of the first term is to ensure
the constraint operator's Hermiticity. As usual, basic operators satisfy commutators
$[\hat{\phi},\hat{p}_{\phi}]=i\hbar$ and $[\hat{b},\hat{\v}]=i\hbar$. As in this method, moments are defined as
\begin{equation}
\G abcd=\langle(\hat{\v}-\v)^a(\hat{b}-b)^b
               (\hat{p}_{\phi}-p_{\phi})^c(\hat{\phi}-\phi)^d\rangle_{\textrm{Weyl}}.
\end{equation}
A special class of semiclassical states, such as Guassion states, allows these moments to have the hierarchy
$\G abcd=O(\hbar^{(a+b+c+d)/2})$. Based on this hierarchy, we can implement a semiclassical expansion,
and for semiclassical approximation, we keep only the moments to a certain order.
From now on, we suppose there exists such a semiclassical state, on which we calculate all the expectation values.

We first consider the non-symmetric constraints $C_{f(q,p)}^{(n)}:=\langle f(\hat{q},\hat{p})\hat{C}^n\rangle$.
In this model, a complete and first class set of quantum constraints can be choosed as
\begin{eqnarray}
&C_Q=\langle\hat{C}\rangle,\quad
C_{\v}=\langle(\hat{\v}-\v)\hat{C}\rangle=\langle\hat{\v}\hat{C}\rangle-\v C_Q,\quad
C_b=\langle(\hat{b}-b)\hat{C}\rangle=\langle\hat{b}\hat{C}\rangle-bC_Q,\quad
\nonumber\\
&C_{p_{\phi}}=\langle(\hat{p}_{\phi}-p_{\phi})\hat{C}\rangle=\langle\hat{p}_{\phi}\hat{C}\rangle-p_{\phi}C_Q,\quad
C_{\phi}=\langle(\hat{\phi}-\phi)\hat{C}\rangle=\langle\hat{\phi}\hat{C}\rangle-\phi C_Q.
\end{eqnarray}
Some of the constraints are not Weyl ordering, but they only differ by some terms of order of $\hbar$.
\begin{eqnarray}
&
(\hat{\v}^2\hat{b}^2)_W
  = \hat{\v}\hat{b}\hat{\v}\hat{b} + i\hbar(\hat{\v}\hat{b})_W + O(\hbar^2),\quad
(\hat{\v}^{-1}\hat{b})_W
  = \hat{b}\hat{\v}^{-1} + \frac12i\hbar\hat{\v}^{-2} + O(\hbar^2),
\nonumber\\
&
(\hat{\v}\hat{b}^3)_W
  = \hat{b}^2\hat{\v}\hat{b} - \frac12i\hbar\hat{b}^2 + O(\hbar^2),\quad
(\hat{p}_{\phi}^2\hat{\phi})_W=\hat{\phi}\hat{p}_{\phi}^2-i\hbar\hat{p}_{\phi}.
\end{eqnarray}
We first re-order the constraints, then Taylor expland them into functions of expectation values and moments.
Considering properties of semiclassical state, for second order we only keep terms to the order of $\hbar$,
we thus have
\begin{eqnarray}
C_Q & = & Ap_{\phi}^2\v^{-1} - B\v b^2 + A\v^{-1}\G0020 - 2Ap_{\phi}\v^{-2}\G1010
          + Ap_{\phi}^2\v^{-3}\G2000 - 2Bb\G1100 - B\v\G0200 ,
\\
\label{equ:chap4:Cv}
C_{\v}
    & = & 2Ap_{\phi}\v^{-1}\G1010 - (Ap_{\phi}^2\v^{-2} + Bb^2)\G2000 - 2B\v b\G1100
          + i\hbar B\v b ,
\\
C_b & = & 2Ap_{\phi}\v^{-1}\G0110 - (Ap_{\phi}^2\v^{-2} + Bb^2)\G1100
          - 2B\v b\G0200 - (\frac12i\hbar Ap_{\phi}^2\v^{-2} + \frac12i\hbar Bb^2) ,
\\
C_{p_{\phi}}
    & = & 2Ap_{\phi}\v^{-1}\G0020 - (Ap_{\phi}^2\v^{-2} + Bb^2)\G1010 - 2B\v b\G0110 ,
\\
C_{\phi}
    & = & 2Ap_{\phi}\v^{-1}\G0011 - (Ap_{\phi}^2\v^{-2} + Bb^2)\G1001 - 2B\v b\G0101 + i\hbar Ap_{\phi}\v^{-1} .
\end{eqnarray}
Here, we can see that $C_{\phi}$ is decoupled from other constraints.

For consistency, the set of all effective constraints should be first class if we start with a single classical constraint
or a first class set of several constraints. Now we check all Poisson brackets between every two effective constraints.
\begin{eqnarray}
&
\{C_Q,C_{\v}\} = - 2BbC_{\v} - 2B\v C_b \approx 0 ,
\quad
\{C_Q,C_{\f}\} = - 2A\v^{-1}C_{\p} + 2A\p\v^{-2}C_{\v} \approx 0 ,
\nonumber\\
&
\{C_Q,C_b\} = 2A\p\v^{-2}C_{\p} - 2B\v^{-1}b^2C_{\v} + 2BbC_b \approx 0 ,
\quad
\{C_Q,C_{\p}\} = 0 ,
\nonumber\\
&
\{C_{\v},C_b\} = - 2A\p\v^{-1}C_{\p} + 4Bb^2C_{\v} + 4B\v b \approx 0 ,
\quad
\{C_{\v},C_{\p}\} = 2B\v bC_{\p} \approx 0 ,
\nonumber\\
&
\{C_{\v},C_{\f}\} = - 2A\p\v^{-1}C_{\v} + 2B\v bC_{\f} \approx 0 ,
\quad
\{C_b,C_{\f}\} = - 2A\p\v^{-1}C_b - 2Bb^2C_{\f} \approx 0 ,
\nonumber\\
&
\{C_b,C_{\p}\} = - 2Bb^2C_{\p} \approx 0 ,
\quad
\{C_{\p},C_{\f}\} = - 4A\p\v^{-1}C_{\p} + 2Bb^2C_{\v} + 2B\v bC_b \approx 0 .
\end{eqnarray}
It shows that this effective constraint system is indeed a first class system.

At this point, a further choice arises: we need to determine which variables we want
to solve in terms of others which are to be kept free. This is related to the choice of time
in a deparametrization procedure. Here, we view $\phi$ as the time variable which is demoted
from a physical variable to the status of an evolution parameter. Notice that time is chosen after
quantization when dealing with effective constraints.

Having made a choice of time, a complete deparametrization requires that
all quantum variables of the form $G^{a_1,b_1}_{a_2,b_2}$ with $a_2 \ne 0$ or $b_2 \ne 0$ be completely constrained or
removed by gauge. Only quantum variables $G^{a,b}_{0,0}$ are allowed to remain free, and must do so
without any further restrictions. To second order, the deparametrized system has $2+3 = 5$
variables; the parametrized theory has $4 + 10 = 14$. We begin by eliminating quantum
variables in favor of the variables associated with the canonical pair $(\phi, p_{\phi})$ only. From the
fact that, on the one hand, $G^{0,0}_{2,0}$, $G^{0,0}_{1,1}$ and $G^{0,0}_{0,2}$ should satisfy the uncertainty relations
and thus cannot all vanish but, on the other hand, are not present in the unconstrained
system, we expect at least one of them to be removed by gauge.
\begin{eqnarray}
\p     & = & A^{-\frac12}B^{\frac12}\v b - \frac12A^{\frac12}B^{-\frac12}\v^{-1}b^{-1}\G0020
             + \v^{-1}\G1010
\nonumber\\
       &   & - \frac12A^{-\frac12}B^{\frac12}\v^{-1}b\G2000 + A^{-\frac12}B^{\frac12}\G1100
             + \frac12A^{-\frac12}B^{\frac12}\v b^{-1}\G0200 ,
\\
\G1010 & = & A^{-\frac12}B^{\frac12}b\G2000 + A^{-\frac12}B^{\frac12}\v\G1100
             - \frac12i\hbar A^{-\frac12}B^{\frac12}\v ,
\\
\G0110 & = & A^{-\frac12}B^{\frac12}b\G1100 + A^{-\frac12}B^{\frac12}\v\G0200
             + \frac12i\hbar A^{-\frac12}B^{\frac12}b ,
\\
\G0020 & = & A^{-1}Bb^2\G2000 + 2 A^{-1}B\v b\G1100 + A^{-1}B\v^2\G0200 ,
\\
\G0011 & = & A^{-\frac12}B^{\frac12}b\G1001 + A^{-\frac12}B^{\frac12}\v\G0101 - \frac12i\hbar .
\end{eqnarray}
The last one is not yet completely expressed in terms of moments only of $(q, p)$. The remaining
moments of $(\phi, p_{\phi})$ are not constrained at all, and thus must be eliminated by gauge
transformations. To summarize, three expectation values are left unconstrained, one of
which should be unphysical; six second-order variables are unconstrained, three of which
should be unphysical. Notice that there is no contradiction to the fact that we have four
weakly commuting (and independent) constraints but expect only three variables to be
removed by gauge. These are constraints on the space of second order moments, which, in
this truncation, as noted before do not have a non-degenerate Poisson bracket (although
the space of all moments has a non-degenerate symplectic structure). Weak commutation
then does not imply first class nature in the traditional sense (see e.g. [\onlinecite{BoSt03}]), and four
weakly commuting constraints may declare less than four variables as gauge. While the
constraints as functionals are independent, their gauge flows may be linearly dependent.

We also emphasize that gauge flows generated by quantum constraints on the quantum
phase space play important roles.
To explicitly account for the unphysical degrees of freedom, we consider the gauge transformations
generated by the constraints. In our method, $C_Q$ is treated as the Hamiltonian constraint
which generates the internal time evolution. Thus in second order, we only consider gauge flows generated
by the four constraints $C_{\p}$, $C_{\v}$, $C_b$ and $C_{\f}$ but not $C_Q$.
After analyzing the gauge transformation, we finally get two observables
\begin{eqnarray}
O_1 & = & \v b + \G1100 - \frac12i\hbar ,
\nonumber\\
O_2 & = & b^2\G2000 + 2\v b\G1100 + \v^2\G0200 .
\end{eqnarray}
And any gauge fixing should not affect these observables. Now we can choose a gauge fixing
\begin{eqnarray}
\G0020 = 0 ,
\quad
\G1001 = \G0101 = 0 ,
\quad
\Longrightarrow
\G0011 = - \frac12i\hbar ,
\end{eqnarray}
and let $\G2000$, $\G1100$ and $\G0200$ to be free as physical variables as well as $\f$ to be free as internal time
parameter.

After substituting the variables required, we can obtain the effective Hamilton constraint,
\begin{equation}
\label{equ:chap4:WDW2ndCQ}
C_Q = Ap_{\phi}^2\v^{-1} - B\v b^2 - 2Bb\G1100 + i\hbar Bb .
\end{equation}
It worth noting that this effective Hamilton constraint $C_Q$ contains an imaginary term.
At the first sight, it seems problematic. But when we think it twice, this manifests the non-commutative property
of quantum operators. It can be expressed into a Moyal *-product inspired by non-commutative geometry.
We can see from the derivation of equation (\ref{equ:chap4:WDW2ndCQ}) that because of using equation (\ref{equ:chap4:Cv}),
it results in an expectation value of a non-symmetric operator, and then leads an imaginary term. In the classical constraint,
it contains a product of $\v$ and $b^2$, but after quantization, they are replaced by the corresponding operators.
Considering the non-commutative property of operators, we can change the usual product into a Moyal *-product,
\begin{equation}
\v*b^2
  = \v e^{\frac{i\hbar}{2}(\overleftarrow{\partial_b}\overrightarrow{\partial_{\v}}
    - \overleftarrow{\partial_{\v}}\overrightarrow{\partial_b})} b^2
  = \v b^2 - i\hbar b .
\end{equation}
Here we can see the imaginary term is coincident with the imaginary term in $C_Q$. So $C_Q$ can be expressed as
\begin{equation}
C_Q = Ap_{\phi}^2\v^{-1} - B\v*b^2 - 2Bb\G1100 .
\end{equation}
This method is correspoding to a procedure of deparameterization of $\phi$,
\begin{equation}
p_{\phi}^2 = A^{-1}B ( \v^2b^2 + 2\v b\G1100 - i\hbar\v b ) .
\end{equation}
Then we can obtain a true Hamiltonian $p_{\phi}=\pm H$ ,
\begin{equation}
H = A^{-\frac12}B^{\frac12} ( \v b + \G1100 - \frac{i\hbar}2 )
  = A^{-\frac12}B^{\frac12} ( \v*b + \G1100 ) .
\end{equation}
The dynamics of the system can also be determined by a struture of Moyal *-product,
\begin{equation}
\dot{f}(b,\v):=\frac1{i\hbar}(f*C_Q-C_Q*f) .
\end{equation}

Since we finally get the effective Hamiltonian constraint $C_Q$,
we can use it to generate the evolution of $p_{\phi}$ and $\phi$,
\begin{equation}
\dot{p_{\phi}}=\{p_{\phi},C_Q\}=0 ,\quad
\dot{\phi}=\{\phi,C_Q\}=2A\v^{-1}p_{\phi} .
\end{equation}
As expected, $\phi$ can be treeted as internal time generating time evolution.
So the time evolution of $\v$ is,
\begin{equation}
\dot{\v} = \{\v,C_Q\} = \frac1{i\hbar}(\v*C_Q-C_Q*\v) = 2B\v b + 2B\G1100 - i\hbar B .
\end{equation}
Then we can calculate Hubble parameter,
\begin{equation}
\label{equ:chap4:hubble}
H_o^2 = \frac{\dot{\v}^2}{9\v^2} = \frac{4B^2}{9} [ b^2 + 2\v^{-1}b\G1100 - i\hbar\v^{-1}b ] .
\end{equation}
In order to eliminate $b$, we use the effective Hamiltonian constraint again. From $C_Q=0$, we can have 
\begin{equation}
b^2 = AB^{-1}\p^2\v^{-2} - 2\v^{-1}b\G1100 + i\hbar\v^{-1}b .
\end{equation}
Put it into Hubble parameter (\ref{equ:chap4:hubble}), keeping terms to order of $\hbar$, we yeild the effective Friedmann equation,
\begin{equation}
H_o^2 = \frac{4B^2}{9} [ AB^{-1}\p^2\v^{-2} ]
      = \frac{8\pi G}{3}\rho .
\end{equation}
where $\rho=\frac{\p^2}{2(2\pi G\gamma\v)^2}$ is the energy density. This is the same as classical cosmology.
There exist a Big Bang singularity.

In the above treatment, the non-symmetric operators in definition of constraints may give rise to imaginary terms
in effective constraints. To avoid this imaginary terms, we may change to choose symmetric constraints.
But one crucial problem is that for this system to be close,
we should find out observables before implementing calculation.
In this system, our goal is to obtain the principle constraint $C_Q$ only with physical variables.
We may not be able to find out enough observables to define constraint at each order
and then to solve for all non-physical variables. But only for the goal we just indicated,
we only need to solve some of the non-physical variables.

The principle constraint is still
\begin{equation}
C_Q = Ap_{\phi}^2\v^{-1} - B\v b^2 + A\v^{-1}\G0020 - 2Ap_{\phi}\v^{-2}\G1010
      + Ap_{\phi}^2\v^{-3}\G2000 - 2Bb\G1100 - B\v\G0200 .
\end{equation}
From the constraint operator (\ref{equ:constr_oper}), we can find that there is only one observable $\hat{\p}$.
Then we can define
\begin{equation}
C_{\p} = \frac12 \langle \hat{\p}\hat{C} + \hat{C}\hat{\p} \rangle - \p C_Q
       = 2A\p\v^{-1}\G0020 - (A\p^2\v^{-2}+Bb^2)\G1010 - 2B\v b\G0110 .
\end{equation}
Obviously one constraint is not enough, and we need to define constraints. We analyze the definition of
these symmetric constraints in detail. And finally we can find that for this constraint system to be close,
we can only include one constraint defined from a non-observable. And we choose
\begin{equation}
C_{\v} = \frac12\langle \hat{\v}\hat{C}+\hat{C}\hat{\v} \rangle - \v C_Q
       = 2A\p\v^{-1}\G1010 - (A\p^2\v^{-2}+Bb^2)\G2000 - 2B\v b\G1100 .
\end{equation}
Now we check the closure of this system.
\begin{eqnarray}
\{ C_{\v},C_{\p} \} 
    & = & \{ \frac12\langle \hat{\v}\hat{C}+\hat{C}\hat{\v} \rangle - \v C_Q ,
             \frac12\langle \hat{\p}\hat{C}+\hat{C}\hat{\p} \rangle - \p C_Q \}
\nonumber\\
    & = & \frac1{4i\hbar}\langle [ \hat{\v}\hat{C}+\hat{C}\hat{\v},
                                   \hat{\p}\hat{C}+\hat{C}\hat{\p} ] \rangle
      =   \frac1{4i\hbar} \langle \hat{\v}\hat{C}[\hat{C},\hat{\p}]
                                  + [\hat{C},\hat{\p}]\hat{C}\hat{\v} \rangle = 0 .
\end{eqnarray}
Also we can have a weak constraint. $\hat{b}\hat{\v}$ and $\hat{\v}\hat{b}$ are weakly commuted with constraint operator,
\begin{equation}
{[\hat{b}\hat{\v},\hat{C}]} = {[\hat{\v}\hat{b},\hat{C}]} = - i\hbar\hat{C} .
\end{equation}
Then we can define another constraint,
\begin{eqnarray}
C_{b\v} & = & \frac12\langle \frac12(\hat{b}\hat{\v}+\hat{\v}\hat{b})\hat{C}
                             + \frac12\hat{C}(\hat{b}\hat{\v}+\hat{\v}\hat{b}) \rangle
              - b\v C_Q - bC_{\v}
\nonumber\\
        & = & 2A\p\G0110 - 2B\v b^2\G1100 - 2B\v^2b\G0200 .
\end{eqnarray}
Now we check whether the Poisson bracket between $C_{\v}$ and $C_{b\v}$ is zero,
\begin{eqnarray}
\{C_{\v},C_{b\v}\} 
    & = & \{ \frac12\langle \hat{\v}\hat{C}+\hat{C}\hat{\v} \rangle - \v C_Q,
             \frac14\langle \hat{b}\hat{\v}\hat{C}+\hat{\v}\hat{b}\hat{C}
                            +\hat{C}\hat{b}\hat{\v}+\hat{C}\hat{\v}\hat{b} \rangle
          - b\v C_Q - bC_{\v} \}
\nonumber\\
    & = & \frac1{8i\hbar}\langle [ \hat{\v}\hat{C}+\hat{C}\hat{\v}, 
                                   \hat{b}\hat{\v}\hat{C}+\hat{\v}\hat{b}\hat{C}
                                   +\hat{C}\hat{b}\hat{\v}+\hat{C}\hat{\v}\hat{b} ] \rangle
\nonumber\\
    & = & \frac1{8i\hbar}
          \langle [\hat{C},\hat{b}\hat{\v}]\hat{C}\hat{\v} + [\hat{C},\hat{\v}\hat{b}]\hat{C}\hat{\v}
                  + \hat{\v}\hat{C}[\hat{C},\hat{b}\hat{\v}] + \hat{\v}\hat{C}[\hat{C},\hat{\v}\hat{b}] \rangle
\nonumber\\
    & = & \frac1{8i\hbar}
          \langle i\hbar\hat{C}^2\hat{\v} + i\hbar\hat{C}^2\hat{\v}
                  + i\hbar\hat{\v}\hat{C}^2 + i\hbar\hat{\v}\hat{C}^2 \rangle = 0 .
\end{eqnarray}
Then we can solve for non-physical observables,
\begin{eqnarray}
\p     & = & A^{-\frac12}B^{\frac12}\v b - \frac12A^{\frac12}B^{-\frac12}\v^{-1}b^{-1}\G0020
             + \v^{-1}\G1010
\nonumber\\
       &   & - \frac12A^{-\frac12}B^{\frac12}\v^{-1}b\G2000 + A^{-\frac12}B^{\frac12}\G1100
             + \frac12A^{-\frac12}B^{\frac12}\v b^{-1}\G0200 ,
\\
\G1010 & = & A^{-\frac12}B^{\frac12}b\G2000 + A^{-\frac12}B^{\frac12}\v\G1100 ,
\\
\G0110 & = & A^{-\frac12}B^{\frac12}b\G1100 + A^{-\frac12}B^{\frac12}\v\G0200 ,
\\
\G0020 & = & A^{-1}Bb^2\G2000 + 2A^{-1}B\v b\G1100 + A^{-1}B\v^2\G0200 .
\end{eqnarray}
After substituting these variables, we have effective Hamiltonian constraint without unphysical variables,
\begin{equation}
C_Q  = A\p^2\v^{-1} - B\v b^2 - 2Bb\G1100 .
\end{equation}
So the time evolution of $\v$ is,
\begin{equation}
\dot{\v} = \{\v,C_Q\} = 2B\v b + 2B\G1100 .
\end{equation}
Then we can calculate Hubble parameter,
\begin{equation}
H_o^2 = \frac{\dot{\v}^2}{9\v^2}
      = \frac{4B^2}{9} [ b^2 + 2\v^{-1}b\G1100 ]
      = \frac{4B^2}{9} [ AB^{-1}\p^2\v^{-2} ]
      = \frac{8\pi G}{3}\rho ,
\end{equation}
where $\rho=\frac{\p^2}{2(2\pi G\gamma\v)^2}$ is again the energy density. Then we come to the same result.

As we know the non-commutative property of operators is a key feature of quantum theory.
Even though now we cast quantum mechanics into classical type of equations,
this key feature is caught in Moyal *-product of non-symmetric constraints
or in order symmetrization of operators in definition of symmetric constraints.

\subsection{Effective Constraints of Loop Quantum Cosmology}
\label{sec:LQC}

For $k=0$ FRW cosmology model, we consider simplified Loop Quantum Cosmology. Its Hamiltonian constraint operator is
\begin{eqnarray}
\hat{C} = - \frac{3}{4\gamma\l^2}\widehat{\sin(2\l b)}\hat{\v}\widehat{\sin(2\l b)}
          + \frac{1}{4\pi G\gamma}\frac{\hat{p}^2}{\hat{\v}}
        = A \hat{p}^2\hat{\v}^{-1} - B \hat{s}\hat{\v}\hat{s} ,
\end{eqnarray}
where $A=\frac1{4\pi G\gamma}$ and $B=\frac3{4\gamma\l^2}$. For convenience,
$\widehat{\sin(2\l b)}$ and $\widehat{\cos(2\l b)}$ are abbreviated as $\hat{s}$ and $\hat{c}$ respectively,
and their expectation values are denoted as $s$ and $c$. For non-symmetric constraints in effective constraint method,
we choose the same constraint functions $C_Q$, $C_{\v}$, $C_b$, $C_{p_{\phi}}$ and $C_{\phi}$ as in WDW quantum cosmology 
\begin{eqnarray}
C_Q    & = & A\p^2\v^{-1} - B\v s^2 + A\v^{-1}\G0020 - 2A\p\v^{-2}\G1010
\nonumber\\
       &   & + A\p^2\v^{-3}\G2000 - 4\l Bsc\G1100 - 4\l^2B\v(c^2-s^2)\G0200 ,
\nonumber\\
C_{\v} & = & 2A\p\v^{-1}\G1010 - (A\p^2\v^{-2}+Bs^2)\G2000 - 4\l B\v sc\G1100 + 2i\hbar\l B\v sc ,
\nonumber\\
C_s    & = & 4\l A\p\v^{-1}c\G0110 - (2\l A\p^2\v^{-2}c+2\l Bs^2c)\G1100 - 8\l^2B\v sc^2\G0200
\nonumber\\
       &   & - i\hbar\l (A\p^2\v^{-2}c+Bs^2c) ,
\nonumber\\
C_{\p} & = & 2A\p\v^{-1}\G0020 - (A\p^2\v^{-2}+Bs^2)\G1010 - 4\l B\v sc\G0110 ,
\nonumber\\
C_{\f} & = & 2A\p\v^{-1}\G0011 - (A\p^2\v^{-2}+Bs^2)\G1001 - 4\l B\v sc\G0101 + i\hbar A\p\v^{-1} .
\end{eqnarray}
We have checked that these constraints form a first class system.
Here $C_{\f}$ is decoupled from other constraints. It can only exert some constriction on unphysical variables $\G0011$, $\G1001$ and $\G0101$. Solving these constraints, we can get
\begin{eqnarray}
\p & = & A^{-\frac12}B^{\frac12}\v s - \frac12A^{\frac12}B^{-\frac12}\v^{-1}s^{-1}\G0020
         + \v^{-1}\G1010
\nonumber\\
   &   & - \frac12A^{-\frac12}B^{\frac12}\v^{-1}s\G2000 + 2\l A^{-\frac12}B^{\frac12}c\G1100
         + 2\l^2A^{-\frac12}B^{\frac12}\v s^{-1}(c^2-s^2)\G0200 ,
\\
\G1010 & = & A^{-\frac12}B^{\frac12}s\G2000 + 2\l A^{-\frac12}B^{\frac12}\v c\G1100
             - i\hbar\l A^{-\frac12}B^{\frac12}\v c ,
\\
\G0110 & = & A^{-\frac12}B^{\frac12}s\G1100 + 2\l A^{-\frac12}B^{\frac12}\v c\G0200
             + \frac12i\hbar A^{-\frac12}B^{\frac12}s ,
\\
\G0020 & = & A^{-1}Bs^2\G2000 + 4\l A^{-1}B\v sc\G1100 + 4\l^2A^{-1}B\v^2c^2\G0200 .
\end{eqnarray}
After analyzing gauge transformation generated by constraints, we can choose gauge fixing
$\G0011=-\frac12i\hbar$, $\G1001=0$ and $\G0101=0$, as well as $\G0002=0$ to saturate
the minimum uncertainty condition $\G0020\G0002-(\G0011)^2=\frac{\hbar^2}4$.

Substituting relevant variables, we can express $C_Q$ in terms of $\G2000$, $\G1100$ and $\G0200$,
obtaining effective Hamiltonian constraint,
\begin{equation}
\label{equ:chap4:LQC2ndCQ}
C_Q = Ap_{\phi}^2\v^{-1} - B\v s^2 - 4\l Bsc\G1100 + 4\l^2B\v s^2\G0200 + 2i\hbar\l Bsc .
\end{equation}
The imaginary term can be understood as the additional term of Moyal *-product.
\begin{equation}
\v*s^2
  = \v e^{\frac{i\hbar}{2}(\overleftarrow{\partial_b}\overrightarrow{\partial_{\v}}
    - \overleftarrow{\partial_{\v}}\overrightarrow{\partial_b})} s^2
  = \v s^2 - 2i\hbar\l\v sc .
\end{equation}
Thus, some terms in $C_Q$ can be expressed as Moyal *-product,
\begin{equation}
C_Q = Ap_{\phi}^2\v^{-1} - B\v*s^2 - 4\l Bsc\G1100 + 4\l^2B\v s^2\G0200 .
\end{equation}
This is corresponding to the procedure of deparameterization,
\begin{equation}
p_{\phi}^2 = A^{-1}B ( \v^2s^2 + 4\l\v sc\G1100 - 4\l^2\v^2s^2\G0200 - 2i\hbar\l\v sc ) .
\end{equation}
Then we can solve for the real Hamiltonian $p_{\phi}=\pm H$,
\begin{equation}
H = A^{-\frac12}B^{\frac12} ( \v*s + 2\l c\G1100 - 2\l^2\v s\G0200 ) .
\end{equation}

As the same way, $\phi$ is treated as internal time, and $C_Q$ can generate time evolution. So the evolution of $\v$ is,
\begin{equation}
\dot{\v} = \{\v,C_Q\} = \frac1{i\hbar}(\v*C_Q-C_Q*\v)
         = 4\l B\v sc + 8\l^2B(c^2-s^2)\G1100 - 16\l^3B\v sc\G0200 - 4i\hbar\l^2B(c^2-s^2) .
\end{equation}
Now we can calculate the Hubble parameter,
\begin{equation}
\label{equ:chap4:LQChubble}
H_o^2 = \frac{\dot{\v}^2}{9\v^2}
      = \frac{16\l^2B^2}{9} [ s^2c^2 + 4\l\v^{-1}sc(c^2-s^2)\G1100 - 8\l^2s^2c^2\G0200
                              - 2i\hbar\l\v^{-1}sc(c^2-s^2) ] .
\end{equation}
In order to eliminate $s$ and $c$, we use again the effective Hamiltonian. From $C_Q=0$, we can get
\begin{eqnarray}
s^2 & = & AB^{-1}p^2\v^{-2} + [ - 4\l\v^{-1}sc\G1100 + 4\l^2s^2\G0200 + 2i\hbar\l\v^{-1}sc ]
      =   AB^{-1}p^2\v^{-2} + O(\hbar) ,
\nonumber\\
c^2 & = & 1 - AB^{-1}p^2\v^{-2} - [ - 4\l\v^{-1}sc\G1100 + 4\l^2s^2\G0200 + 2i\hbar\l\v^{-1}sc ] ,
\nonumber\\
s^2c^2 & = & AB^{-1}p^2\v^{-2}(1-AB^{-1}p^2\v^{-2}) + (1-2AB^{-1}p^2\v^{-2})[ - 4\l\v^{-1}sc\G1100
\nonumber\\
       &   & + 4\l^2s^2\G0200 + 2i\hbar\l\v^{-1}sc ]
         =   AB^{-1}p^2\v^{-2}(1-AB^{-1}p^2\v^{-2}) + O(\hbar) ,
\nonumber\\
sc     & = & \sqrt{AB^{-1}p^2\v^{-2}(1-AB^{-1}p^2\v^{-2})} + O(\hbar) ,
\nonumber\\
c^2-s^2
       & = & 1 - 2AB^{-1}p^2\v^{-2} + O(\hbar) .
\end{eqnarray}
Put them into Hubble parameter (\ref{equ:chap4:LQChubble}), and keep terms to order of $\hbar$,
we can obtain modified Friedmann equation,
\begin{eqnarray}
H_o^2 & = & \frac{16\l^2B^2}{9} [ AB^{-1}p^2\v^{-2}(1-AB^{-1}p^2\v^{-2})
                                  - 4\l^2AB^{-1}p^2\v^{-2}\G0200 ]
\nonumber\\
      & = & \frac{8\pi G}{3}\rho
            \Big[ 1 - \frac1{\rho_c} \Big( \rho + \frac3{2\pi G\gamma^2}\G0200 \Big) \Big] ,
\end{eqnarray}
where, $\rho=\frac{p^2}{2(2\pi G\gamma\v)^2}$ is the energy density,
and $\rho_c=\frac{3}{8\pi G\gamma^2\l^2}$ is the critical energy density.

Now we can see, when $\rho=\rho_c-\frac3{2\pi G\gamma^2}\G0200$, Hubble parameter reaches zero.
Suppose today's cosmology can be described by a semi-classical state,
as we evolve this state backward in the internal time,
the scale of cosmology keeps shrinking. However, we will not touch the classical singularity.
Before we reach that point, we first meet the situation of $\rho=\rho_c-\frac3{2\pi G\gamma^2}\G0200$.
And at this point the Hubble parameter vanishes, the scale of cosmology reaches its minimum and then starts to become larger.
It is corresponding to a bounce generated by quantum effects of Loop Quantum Cosmology.
In this sense, we solve the problem of singulary, that is to say the Big Bang singulary is replaced by this quantum bounce.
This result is different from the former one in other methods[\onlinecite{AsPa06-1,AsPa06-2,Ta08,DiMa08}]
only by a term $\G0200$ in critical energy density, which is of the order of $\hbar$ and we can just neglect it.

\subsection{3rd order effective constraints of quantum cosmology}

So far, we only consider momenta up to order of $hbar$, that is the terms $\G abcd,a+b+c+d=2$.
And now we consider the third order, which are of order of $\hbar^{\frac32}$, $\G abcd,a+b+c+d=3$.
It emerges something new from this higer order system.

In WDW cosmology, the principle constraint is (for convenience, we simplify $\hat{p_{\phi}}$ as $\hat{p}$, and $p_{\phi}$ as $p$)
\begin{eqnarray}
C_Q & = & \langle A \hat{p}^2\hat{\v}^{-1} - B \hat{\v}\hat{b}^2 \rangle_W
\nonumber\\
    & = & Ap^2\v^{-1} - B\v b^2 + A\v^{-1}\G0020 - 2Ap\v^{-2}\G1010 + Ap^2\v^{-3}\G2000 - 2Bb\G1100 - B\v\G0200
\nonumber\\
    &   & - A\v^{-2}\G1020 + 2Ap\v^{-3}\G2010 - Ap^2\v^{-4}\G3000 - B\G1200
\end{eqnarray}
From $C_Q=0$, we can get $p = A^{-\frac12}B^{\frac12}\v b + O(\hbar)$
So, all constraints become,
\begin{eqnarray}
C_Q & = & Ap^2\v^{-1} - B\v b^2 + A\v^{-1}\G0020 - 2A^{\frac12}B^{\frac12}\v^{-1}b\G1010 + B\v^{-1}b^2\G2000 - 2Bb\G1100
\nonumber\\
    &   & - B\v\G0200 - A\v^{-2}\G1020 + 2A^{\frac12}B^{\frac12}\v^{-2}b\G2010 - B\v^{-2}b^2\G3000
          - B\G1200 ,
\end{eqnarray}
\begin{eqnarray}
C_{\v} & = & 2A^{\frac12}B^{\frac12}b\G1010 - 2Bb^2\G2000 - 2B\v b\G1100 + i\hbar B\v b
\nonumber\\
       &   & + A\v^{-1}\G1020 - 2A^{\frac12}B^{\frac12}\v^{-1}b\G2010 + B\v^{-1}b^2\G3000
             - 2Bb\G2100 - B\v\G1200 ,
\\
C_b    & = & 2A^{\frac12}B^{\frac12}b\G0110 - 2Bb^2\G1100 - 2B\v b\G0200 - i\hbar Bb^2
\nonumber\\
       &   & + A\v^{-1}\G0120 - 2A^{\frac12}B^{\frac12}\v^{-1}b\G1110 + B\v^{-1}b^2\G2100
             - 2Bb\G1200 - B\v\G0300 ,
\\
C_p    & = & 2A^{\frac12}B^{\frac12}b\G0020 - 2Bb^2\G1010 - 2B\v b\G0110
\nonumber\\
       &   & + A\v^{-1}\G0030 - 2A^{\frac12}B^{\frac12}\v^{-1}b\G1020 + B\v^{-1}b^2\G2010
             - 2Bb\G1110 - B\v\G0210 ,
\\
C_{\v^2}  & = & \langle A \hat{p}^2\hat{\v}
                        - B [ \hat{\v}^3\hat{b}^2 - 2i\hbar\hat{\v}^2\hat{b} ] \rangle_W
                - 2\v C_{\v} - \v^2C_Q
\nonumber\\
          & = & 2A^{\frac12}B^{\frac12}b\G2010 - 2Bb^2\G3000 - 2B\v b\G2100 ,
\\
C_{b^2}   & = & \langle A [ \hat{p}^2\hat{\v}^{-1}\hat{b}^2 - i\hbar\hat{p}^2\hat{\v}^{-2}\hat{b} ]
                        - B [ \hat{\v}\hat{b}^4 + i\hbar\hat{b}^3 ] \rangle_W - 2bC_b - b^2C_Q
\nonumber\\
          & = & 2A^{\frac12}B^{\frac12}b\G0210 - 2Bb^2\G1200 - 2B\v b\G0300 ,
\\
C_{p^2}   & = & \langle A \hat{p}^4\hat{\v}^{-1} - B \hat{p}^2\hat{\v}\hat{b}^2 \rangle_W
                - 2pC_p - p^2C_Q
\nonumber\\
          & = & 2A^{\frac12}B^{\frac12}b\G0030 - 2Bb^2\G1020 - 2B\v b\G0120 ,
\\
\label{equ:chap4:WDW3rdCbv}
C'_{b\v}  & = & \langle A \hat{p}^2\hat{b} - B \hat{\v}^2\hat{b}^3 \rangle_W
                - bC_{\v} - \v C_b - b\v C_Q
\nonumber\\
          & = & 2A^{\frac12}B^{\frac12}b\G1110 - 2Bb^2\G2100 - 2B\v b\G1200 ,
\\
C'_{\v b} & = & \langle A \hat{p}^2\hat{b} - B [ \hat{\v}^2\hat{b}^3 - i\hbar\v b^2 ] \rangle_W
                - \v C_b - bC_{\v} - \v bC_Q
\nonumber\\
          & = & i\hbar B\v b^2 + 2A^{\frac12}B^{\frac12}b\G1110 - 2Bb^2\G2100 - 2B\v b\G1200 ,
\\
C_{b\v}   & = & \frac12 ( C'_{b\v} + C'_{\v b} )
            =   \frac12i\hbar B\v b^2 + 2A^{\frac12}B^{\frac12}b\G1110 - 2Bb^2\G2100
                - 2B\v b\G1200 ,
\\
C_{p\v}   & = & \langle A \hat{p}^3
                        - B [ \hat{p}\hat{\v}^2\hat{b}^2 - i\hbar\hat{p}\hat{\v}\hat{b} ] \rangle_W
                - pC_{\v} - {\v}C_p - p\v C_Q
\nonumber\\
          & = & 2A^{\frac12}B^{\frac12}b\G1020 - 2Bb^2\G2010 - 2B\v b\G1110 ,
\\
C_{pb}    & = & \langle A [ \hat{p}^3\hat{\v}^{-1}\hat{b} - \frac12i\hbar\hat{p}^3\hat{\v}^{-2} ]
                        - B [ \hat{p}\hat{\v}\hat{b}^3 + \frac12i\hbar\hat{p}\hat{b}^2 ] \rangle_W
                - pC_b - bC_p - pbC_Q
\nonumber\\
          & = & 2A^{\frac12}B^{\frac12}b\G0120 - 2Bb^2\G1110 - 2B\v b\G0210.
\end{eqnarray}

Solving for momenta, we have
\begin{eqnarray}
\G1110 & = & - \frac14i\hbar A^{-\frac12}B^{\frac12}\v b + A^{-\frac12}B^{\frac12}b\G2100
             + A^{-\frac12}B^{\frac12}\v \G1200 ,
\\
\G2010 & = & A^{-\frac12}B^{\frac12}b\G3000 + A^{-\frac12}B^{\frac12}\v\G2100 ,
\\
\G0210 & = & A^{-\frac12}B^{\frac12}b\G1200 + A^{-\frac12}B^{\frac12}\v\G0300 ,
\\
\G1020 & = & - \frac14i\hbar A^{-1}B\v^2b + A^{-1}Bb^2\G3000 + 2A^{-1}B\v b\G2100
             + A^{-1}B\v^2\G1200 ,
\\
\G0120 & = & - \frac14i\hbar A^{-1}B\v b^2 + A^{-1}Bb^2\G2100 + 2A^{-1}B\v b\G1200
             + A^{-1}B\v^2\G0300 ,
\\
\G0030 & = & - \frac12i\hbar A^{-\frac32}B^{\frac32}\v^2b^2 + A^{-\frac32}B^{\frac32}b^3\G3000
\nonumber\\
       &   & + 3A^{-\frac32}B^{\frac32}\v b^2\G2100 + 3A^{-\frac32}B^{\frac32}\v^2b\G1200
             + A^{-\frac32}B^{\frac32}\v^3\G0300 ,
\\
\G1010 & = & A^{-\frac12}B^{\frac12}b\G2000 + A^{-\frac12}B^{\frac12}\v\G1100
             - \frac38i\hbar A^{-\frac12}B^{\frac12}\v + A^{-\frac12}B^{\frac12}\G2100 ,
\\
\G0110 & = & A^{-\frac12}B^{\frac12}b\G1100 + A^{-\frac12}B^{\frac12}\v\G0200
             + \frac38i\hbar A^{-\frac12}B^{\frac12}b + A^{-\frac12}B^{\frac12}\G1200 ,
\\
\G0020 & = & A^{-1}Bb^2\G2000 + 2A^{-1}B\v b\G1100 + A^{-1}B\v^2\G0200 - \frac14i\hbar A^{-1}B\v b
\nonumber\\
       &   & + 2A^{-1}Bb\G2100 + 2A^{-1}B\v\G1200 .
\end{eqnarray}
Substituting all relevant quantities into principle constraints and expressing it into a function of quantum variables, finally we obtain the effective Hamiltonian constraint,
\begin{equation}
C_Q = Ap^2\v^{-1} - B\v b^2 - 2Bb\G1100 + \frac34i\hbar Bb .
\end{equation}
Though we take third order momenta into acount, the resulting effective Hamiltonian acquiring no higher order correction. That is to say, for WDW Quantum Cosmology, there is no third order correction in effective Hamiltonian constraint when considering quantum effects. Also we can see that comparing with the one (\ref{equ:chap4:WDW2ndCQ}) in the second order situation, it is only different by a term $-\frac14i\hbar Bb$. It comes from an effect of the order of quantum operators in $C_{b\v}$. If we choose $C'_{b\v}$, i.e. equation (\ref{equ:chap4:WDW3rdCbv}) as constraint instead of $C_{b\v}$, then the effective Hamiltonian constraint is totally the same as the one in second order,
\begin{equation}
C_Q = Ap^2\v^{-1} - B\v b^2 - 2Bb\G1100 + i\hbar Bb .
\end{equation}

But this will not affect the Hubble parameter and effective Friedmann equation,
\begin{eqnarray}
\dot{\v} & = & \{\v,C_Q\} = 2B\v b + 2B\G1100 - \frac34i\hbar B ,
\\
H_o^2    & = & \frac{\dot{\v}^2}{9\v^2}
           =   \frac{4B^2}{9} [ b^2 + 2\v^{-1}b\G1100 - \frac34i\hbar\v^{-1}b ] .
\end{eqnarray}
Using $C_Q=0$, we can derive $b^2 = AB^{-1}p^2\v^{-2} - 2\v^{-1}b\G1100 + \frac34i\hbar\v^{-1}b$, then
\begin{equation}
H_o^2 = \frac{4B^2}{9} [ AB^{-1}p^2\v^{-2} ]
      = \frac{8\pi G}{3}\rho ,
\end{equation}
where $\rho=\frac{p^2}{2(2\pi G\gamma\v)^2}$ is still the energy density. This is exactly the same as in second order.

Now we continue to see the situation in Loop Quantum Cosmology, its principle constraint is,
\begin{eqnarray}
C_Q    & = & \langle A \hat{p}^2\hat{\v}^{-1} - B \hat{\v}\hat{s}^2 \rangle_W
\nonumber\\
       & = & Ap^2\v^{-1} - B\v s^2 + A\v^{-1}\G0020 - 2Ap\v^{-2}\G1010 + Ap^2\v^{-3}\G2000
\nonumber\\
       &   & - 4\l Bsc\G1100 - 4\l^2B\v(c^2-s^2)\G0200 - A\v^{-2}\G1020 + 2Ap\v^{-3}\G2010
\nonumber\\
       &   & - Ap^2\v^{-4}\G3000 - 4\l^2B(c^2-s^2)\G1200 + \frac{32}3\l^3B\v sc\G0300 .
\end{eqnarray}
From $C_Q=0$, we can get $p = A^{-\frac12}B^{\frac12}\v s + O(\hbar)$. And constraints become
\begin{eqnarray}
C_Q    & = & Ap^2\v^{-1} - B\v s^2 + A\v^{-1}\G0020 - 2A^{\frac12}B^{\frac12}\v^{-1}s\G1010
             + B\v^{-1}s^2\G2000
\nonumber\\
       &   & - 4\l Bsc\G1100 - 4\l^2B\v(c^2-s^2)\G0200 - A\v^{-2}\G1020
             + 2A^{\frac12}B^{\frac12}\v^{-2}s\G2010
\nonumber\\
       &   & - B\v^{-2}s^2\G3000 - 4\l^2B(c^2-s^2)\G1200 + \frac{32}3\l^3B\v sc\G0300 ,
\\
C_{\v} & = & 2A^{\frac12}B^{\frac12}s\G1010 - 2Bs^2\G2000 - 4\l B\v sc\G1100 + 2i\hbar\l B\v sc
             + A\v^{-1}\G1020
\nonumber\\
       &   & - 2A^{\frac12}B^{\frac12}\v^{-1}s\G2010 + B\v^{-1}s^2\G3000 - 4\l Bsc\G2100
             - 4\l^2B\v(c^2-s^2)\G1200 ,
\\
C_s    & = & 4\l A^{\frac12}B^{\frac12}sc\G0110 - 4\l Bs^2c\G1100 - 8\l^2B\v sc^2\G0200
             - 2i\hbar\l Bs^2c
\nonumber\\
       &   & + 2\l A\v^{-1}c\G0120 - 4\l A^{\frac12}B^{\frac12}\v^{-1}sc\G1110
             - 4\l^2A^{\frac12}B^{\frac12}s^2\G0210
\nonumber\\
       &   & + 2\l B\v^{-1}s^2c\G2100 - 4\l^2Bs(2c^2-s^2)\G1200 - 8\l^3B\v c(c^2-2s^2)\G0300 ,
\\
C_p    & = & 2A^{\frac12}B^{\frac12}s\G0020 - 2Bs^2\G1010 - 4\l B\v sc\G0110 + A\v^{-1}\G0030
\nonumber\\
       &   & - 2A^{\frac12}B^{\frac12}\v^{-1}s\G1020 + B\v^{-1}s^2\G2010 - 4\l Bsc\G1110
             - 4\l^2B\v(c^2-s^2)\G0210 ,
\\
C_{\v^2}  & = & \langle A \hat{p}^2\hat{\v}
                        - B [ \hat{\v}^3\hat{s}^2 - 4i\hbar\l\hat{\v}^2\hat{s}\hat{c}] \rangle_W
                - 2\v C_{\v} - \v^2C_Q
\nonumber\\
          & = & 2A^{\frac12}B^{\frac12}s\G2010 - 2Bs^2\G3000 - 4\l B\v sc\G2100 ,
\end{eqnarray}
\begin{eqnarray}
C_{s^2}   & = & \langle A [ \hat{p}^2\hat{\v}^{-1}\hat{s}^2
                            - 2i\hbar\l\hat{p}^2\hat{\v}^{-2}\hat{s}\hat{c} ]
                        - B [ \hat{\v}\hat{s}^4 + 2i\hbar\l\hat{s}^3\hat{c} ] \rangle_W
                - 2sC_s - s^2C_Q
\nonumber\\
          & = & 8\l^2A^{\frac12}B^{\frac12}sc^2\G0210 - 8\l^2Bs^2c^2\G1200 - 16\l^3B\v sc^3\G0300 ,
\\
C_{p^2}   & = & \langle A \hat{p}^4\hat{\v}^{-1} - B \hat{p}^2\hat{\v}\hat{s}^2 \rangle_W
                - 2pC_p - p^2C_Q
\nonumber\\
          & = & 2A^{\frac12}B^{\frac12}s\G0030 - 2Bs^2\G1020 - 4\l B\v sc\G0120 ,
\\
\label{equ:chap4:LQC3rdCbv}
C'_{s\v}  & = & \langle A \hat{p}^2\hat{s} - B \hat{\v}^2\hat{s}^3 \rangle_W
                - sC_{\v} - \v C_s - s\v C_Q
\nonumber\\
          & = & 4\l A^{\frac12}B^{\frac12}sc\G1110 - 4\l Bs^2c\G2100 - 8\l^2B\v sc^2\G1200 ,
\\
C'_{\v s} & = & \langle A \hat{p}^2\hat{s} - B [ \hat{\v}^2\hat{s}^3 - 2i\hbar\l\v s^2c ] \rangle_W
                - \v C_s - sC_{\v} - \v sC_Q
\nonumber\\
          & = & 2i\hbar\l B\v s^2c + 4\l A^{\frac12}B^{\frac12}sc\G1110 - 4\l Bs^2c\G2100
                - 8\l^2B\v sc^2\G1200 ,
\\
C_{s\v}   & = & \frac12 ( C'_{s\v} + C'_{\v s} )
            =   i\hbar\l B\v s^2c + 4\l A^{\frac12}B^{\frac12}sc\G1110 - 4\l Bs^2c\G2100
                - 8\l^2B\v sc^2\G1200 ,
\\
C_{p\v}   & = & \langle A \hat{p}^3
                        - B [ \hat{p}\hat{\v}^2\hat{s}^2
                              - 2i\hbar\l\hat{p}\hat{\v}\hat{s}\hat{c}] \rangle_W
                - pC_{\v} - {\v}C_p - p\v C_Q
\nonumber\\
          & = & 2A^{\frac12}B^{\frac12}s\G1020 - 2Bs^2\G2010 - 4\l B\v sc\G1110 ,
\\
C_{ps}    & = & \langle A [ \hat{p}^3\hat{\v}^{-1}\hat{s} - i\hbar\l\hat{p}^3\hat{\v}^{-2}\hat{c} ]
                        - B [ \hat{p}\hat{\v}\hat{s}^3
                              + i\hbar\l\hat{p}\hat{s}^2\hat{c} ] \rangle_W
                - pC_s - sC_p - psC_Q
\nonumber\\
          & = & 4\l A^{\frac12}B^{\frac12}sc\G0120 - 4\l Bs^2c\G1110 - 8\l^2B\v sc^2\G0210 .
\end{eqnarray}

Solved all the constraints, we have momenta
\begin{eqnarray}
\G1110 & = & - \frac14i\hbar A^{-\frac12}B^{\frac12}\v s + A^{-\frac12}B^{\frac12}s\G2100
             + 2\l A^{-\frac12}B^{\frac12}\v c\G1200 ,
\\
\G2010 & = & A^{-\frac12}B^{\frac12}s\G3000 + 2\l A^{-\frac12}B^{\frac12}\v c\G2100 ,
\\
\G0210 & = & A^{-\frac12}B^{\frac12}s\G1200 + 2\l A^{-\frac12}B^{\frac12}\v c\G0300 ,
\\
\G1020 & = & - \frac12i\hbar\l A^{-1}B\v^2sc + A^{-1}Bs^2\G3000 + 4\l A^{-1}B\v sc\G2100
             + 4\l^2A^{-1}B\v^2c^2\G1200 ,
\\
\G0120 & = & - \frac14i\hbar A^{-1}B\v s^2 + A^{-1}Bs^2\G2100 + 4\l A^{-1}B\v sc\G1200
             + 4\l^2A^{-1}B\v^2c^2\G0300 ,
\\
\G0030 & = & - i\hbar\l A^{-\frac32}B^{\frac32}\v^2s^2c + A^{-\frac32}B^{\frac32}s^3\G3000
             + 6\l A^{-\frac32}B^{\frac32}\v s^2c\G2100 ,
\nonumber\\
       &   & + 12\l^2A^{-\frac32}B^{\frac32}\v^2sc^2\G1200
             + 8\l^3A^{-\frac32}B^{\frac32}\v^3c^3\G0300 ,
\\
\G1010 & = & A^{-\frac12}B^{\frac12}s\G2000 + 2\l A^{-\frac12}B^{\frac12}\v c\G1100
             - \frac34i\hbar\l A^{-\frac12}B^{\frac12}\v c
\nonumber\\
       &   & + 2\l A^{-\frac12}B^{\frac12}c\G2100 - 2\l^2A^{-\frac12}B^{\frac12}\v s\G1200 ,
\\
\G0110 & = & A^{-\frac12}B^{\frac12}s\G1100 + 2\l A^{-\frac12}B^{\frac12}\v c\G0200
             + \frac38i\hbar A^{-\frac12}B^{\frac12}s
\nonumber\\
       &   & + 2\l A^{-\frac12}B^{\frac12}c\G1200 - 2\l^2A^{-\frac12}B^{\frac12}\v s\G0300 ,
\\
\G0020 & = & A^{-1}Bs^2\G2000 + 4\l A^{-1}B\v sc\G1100 + 4\l^2A^{-1}B\v^2c^2\G0200
             - \frac12i\hbar\l A^{-1}B\v sc
\nonumber\\
       &   & + 4\l A^{-1}Bsc\G2100 + 4\l^2A^{-1}B\v(2c^2-s^2)\G1200 - 8\l^3A^{-1}B\v^2sc\G0300 .
\end{eqnarray}

Substituting terms into principle constraint, we finally obtain the effective constraint,
\begin{eqnarray}
C_Q    & = & Ap^2\v^{-1} - B\v s^2 - 4\l Bsc\G1100 + 4\l^2B\v s^2\G0200 + \frac32i\hbar\l Bsc
\nonumber\\
       &   & + 4\l^2Bs^2\G1200 + \frac83\l^3B\v sc\G0300 .
\end{eqnarray}
Here we can see the above equation acquires third order momenta correction. In other words, there are higher order corrections in Loop Quantum Cosmology when considering quantum effects. The imaginary term here is different from the one in second order (\ref{equ:chap4:LQC2ndCQ}). This is also due to the uncertainty of the order of operators in higher order constraint $C_{s\v}$. If we use $C'_{s\v}$, i.e. equation (\ref{equ:chap4:LQC3rdCbv}), then this imaginary term comes back to the one in second order situation. The higher order corrections emerging from the effective Hamiltonian constraint make crutial differences in Hubble parameter and effective Friedmann equation comparing with ones in WDW Quantum Cosmology,
\begin{eqnarray}
\dot{\v} & = & \{\v,C_Q\} = 4\l B\v sc + 8\l^2B(c^2-s^2)\G1100 - 16\l^3B\v sc\G0200
               - 3i\hbar\l^2B(c^2-s^2)
\nonumber\\
         &   & - 16\l^3Bsc\G1200 - \frac{16}3\l^4B\v(c^2-s^2)\G0300 ,
\\
H_o^2    & = & \frac{\dot{\v}^2}{9\v^2}
           =   \frac{16\l^2B^2}{9}[s^2c^2 + 4\l\v^{-1}sc(c^2-s^2)\G1100 - 8\l^2s^2c^2\G0200
\nonumber\\
         &   & - \frac32i\hbar\l\v^{-1}sc(c^2-s^2) - 8\l^2\v^{-1}s^2c^2\G1200
               - \frac83\l^3sc(c^2-s^2)\G0300] .
\end{eqnarray}
In order to eliminate $s$ and $c$ in above equation, we use again the effective Hamiltonian constraint. From $C_Q=0$, we have
\begin{eqnarray}
s^2     & = & AB^{-1}p^2\v^{-2} + [ - 4\l\v^{-1}sc\G1100 + 4\l^2s^2\G0200
              + \frac32i\hbar\l\v^{-1}sc
\nonumber\\
        &   & + 4\l^2\v^{-1}s^2\G1200 + \frac83\l^3sc\G0300 ] = AB^{-1}p^2\v^{-2} + O(\hbar) ,
\nonumber\\
c^2     & = & 1 - AB^{-1}p^2\v^{-2} - [ - 4\l\v^{-1}sc\G1100 + 4\l^2s^2\G0200
              + \frac32i\hbar\l\v^{-1}sc
\nonumber\\
        &   & + 4\l^2\v^{-1}s^2\G1200 + \frac83\l^3sc\G0300 ] = 1 - AB^{-1}p^2\v^{-2} + O(\hbar) ,
\nonumber\\
s^2c^2  & = & AB^{-1}p^2\v^{-2}(1-AB^{-1}p^2\v^{-2}) + (1-2AB^{-1}p^2\v^{-2})[ - 4\l\v^{-1}sc\G1100
              + 4\l^2s^2\G0200
\nonumber\\
        &   & + \frac32i\hbar\l\v^{-1}sc + 4\l^2\v^{-1}s^2\G1200 + \frac83\l^3sc\G0300 ]
          =   AB^{-1}p^2\v^{-2}(1-AB^{-1}p^2\v^{-2}) + O(\hbar) ,
\nonumber\\
sc      & = & \sqrt{AB^{-1}p^2\v^{-2}(1-AB^{-1}p^2\v^{-2})} + O(\hbar) ,
\nonumber\\
c^2-s^2 & = & 1 - 2AB^{-1}p^2\v^{-2} + O(\hbar) .
\end{eqnarray}
Substitute them into Hubble parameter, we can have the modified effective Friedmann equation,
\begin{eqnarray}
H_o^2 & = & \frac{16\l^2B^2}{9}AB^{-1}p^2\v^{-2}[1-AB^{-1}p^2\v^{-2} - 4\l^2\G0200
            - 4\l^2\v^{-1}\G1200]
\nonumber\\
      & = & \frac{8\pi G}{3}\rho
            \Big[ 1 - \frac1{\rho_c}
                  \Big( \rho + \frac3{2\pi G\gamma^2}
                        \big(\G0200 - \v^{-1}\G1200
                        \big)
                  \Big)
            \Big] ,
\end{eqnarray}
where $\rho=\frac{p^2}{2(2\pi G\gamma\v)^2}$ is the energy density, while $\rho_c=\frac{3}{8\pi G\gamma^2\l^2}$ is the critical energy density. The result shows that there is an additional momentum $\v^{-1}\G1200$ with the order of $\hbar^{\frac32}$.

\section{Discussion}\label{sec:discussion}

In the above discussion, we used the method of effective constraints to obtain the effective Hamiltonian constraints both in WDW Quantum Cosmology and Loop Quantum Cosmology, and then used them to analyze the dynamics of those systems. In this method, we assumed that there exist physical semi-classical state, and based on this state we calculated all the expectation values and expanded them into functions of expectation values and momenta of basic operators.

Since this state is general, this method should be applicable to a wild range of situations. The analyzing of the case of a specific type of semi-classical state, coherent state, has convinced us its reliability. Now we compare our result with the one from the method of path integral using a coherent state to obtain the effective Hamiltonian [\onlinecite{ACH10,QHM10,QDM12,QM12}]. The resulting effective Hamiltonian in the reference [\onlinecite{QM12}] is, (considering the difference in denotations)
\begin{eqnarray}
\mathscr{H}_{\textrm{eff}}
  & = & - p_{\phi}^2 - \frac{\hbar^2}{2\sigma^2} + 12\pi G
        \Big[ \Big( \v^2+\frac{\delta^2}2 \Big)
              \Big( b^2+\frac{\hbar^2}{2\delta^2} \Big)
              -i\hbar\v b \Big]
\nonumber\\
  & = & - p_{\phi}^2 - \frac{\hbar^2}{2\sigma^2} + 12\pi G
        \Big[ \v*(\v b^2) + \frac{\delta^2}2b^2 + \frac{\hbar^2}{2\delta^2}\v^2
              + \frac{\hbar^2}4 \Big] .
\end{eqnarray}
It contains the fluctuation $\frac{\hbar^2}{2\sigma^2}$ of $p_{\phi}$ corresponding to $\G0020$, the fluctuation $\frac{\delta^2}2$ of $\v$ corresponding to $\G2000$ and the fluctuation $\frac{\hbar^2}{2\delta^2}$ of $b$ which is the same as $\G0200$. Since they choose the coherent state as the semi-classical state, the corelation function between $b$ and $\v$ is itself vanishing. Thus there is no such term appearing in the effective Hamiltonian. In this effective Hamiltonian constraint it contians a fluctuation term of $p_{\phi}$, while in our method of effective constraints, we have solved the corresponding term $\G0020$ and put it in terms of other functions. If we keep this term, then the effective Hamiltonian constraint becoms,
\begin{equation}
C'_Q = Ap_{\phi}^2\v^{-1} - B\v b^2 + A\v^{-1}\G0020 - B\v^{-1}b^2\G2000 - 4Bb\G1100
       - B\v\G0200 + i\hbar Bb .
\end{equation}
Then we can obtain
\begin{eqnarray}
\mathscr{H}'_{\textrm{eff}}
  & = & - p_{\phi}^2 - \G0020 + 12\pi G
        \Big( \v^2b^2 + b^2\G2000 + 4\v b\G1100 + \v^2\G0200 - i\hbar\v b \Big)
\nonumber\\
  & = & - p_{\phi}^2 - \G0020 + 12\pi G
        \Big( \v*(\v b^2) + b^2\G2000 + 4\v b\G1100 + \v^2\G0200 \Big) .
\end{eqnarray}
From above equations, we can see these two results match to each other without higher order terms.

For comparison, we change the variables both in WDW Quantum Cosmology and Loop Quantum Cosmology,
\begin{equation}
b:=\frac{\sqrt{\Delta}c}{2|p|^{\frac12}} ,\quad
\nu:=\frac{\textrm{sgn}(p)|p|^{\frac32}}{2\pi G\gamma\sqrt{\Delta}} ,
\end{equation}
where they satisfy the basic Poisson bracket $\{b,\v\}=1$. In Loop Quantum Cosmology, the constraint operator is,
\begin{eqnarray}
\hat{C} = - \frac{3}{4\gamma}\widehat{\sin(2b)}\hat{\v}\widehat{\sin(2b)}
          + \frac{1}{4\pi G\gamma}\frac{\hat{p}^2}{\hat{\v}}
        = A \hat{p}^2\hat{\v}^{-1} - B \hat{s}\hat{\v}\hat{s} ,
\end{eqnarray}
where $A=\frac{1}{4\pi G\gamma}$, $B=\frac{3}{4\gamma}$. In the paper[\onlinecite{QM12}], they use the coherence state to get the Hamiltonian,
\begin{eqnarray}
\underline{\mathscr{H}}_{\textrm{eff}}
  & = & - p_{\phi}^2 - \frac{\hbar^2}{2\sigma^2} + 3\pi G
      \Big[ \Big(\v^2+\frac{d^2}2\Big)
            \Big(\sin^2(2b)(1-\frac{4\hbar^2}{d^2})+\frac{2\hbar^2}{d^2}\Big)
            - i\hbar\v\sin(4b)(1-\frac{4\hbar^2}{d^2}) \Big]
\nonumber\\
  & = & - p_{\phi}^2 - \frac{\hbar^2}{2\sigma^2} + 3\pi G
      \Big[ \v*(\v\sin^2(2b)(1-\frac{4\hbar^2}{d^2})) + \frac{2\hbar^2}{d^2}\v^2
            + \frac{d^2}2\sin^2(2b)(1-\frac{4\hbar^2}{d^2}) + \hbar^2
      \Big] ,
\nonumber\\
\end{eqnarray}
where it contains the fluctuation $\frac{\hbar^2}{2\sigma^2}$ of $p_{\phi}$ corresponding to $\G0020$, the fluction $\frac{d^2}2$ of $\v$ corresponding to $\G2000$ and the fluctuation $\frac{2\hbar^2}{d^2}$ of $b$ which is the same as $\G0200$. Additionally it includes a higher order correction term $\frac{4\hbar^2}{d^2}$ of the variable $\sin^2(2b)$. We suppose that this term is ignorable, then
\begin{eqnarray}
\underline{\mathscr{H}}_{\textrm{eff}}
  & = & - p_{\phi}^2 - \frac{\hbar^2}{2\sigma^2} + 3\pi G
        \Big[ \Big(\v^2+\frac{d^2}2\Big)
              \Big(\sin^2(2b)+\frac{2\hbar^2}{d^2}\Big)
              - i\hbar\v\sin(4b) \Big]
\nonumber\\
\label{equ:chap4:CohMeth}
  & = & - p_{\phi}^2 - \frac{\hbar^2}{2\sigma^2} + 3\pi G
        \Big[ \v*(\v\sin^2(2b)) + \frac{d^2}2\sin^2(2b) + \frac{2\hbar^2}{d^2}\v^2 + \hbar^2
        \Big] .
\end{eqnarray}
Because they use the coherence state as the semi-classical state, the corelation function between $b$ and $\v$ is itself vanishing. In the method we introduced, keeping $\G0200$, the hamiltonian constraint is,
\begin{eqnarray}
\label{equ:chap4:EffConMeth}
C'_Q & = & Ap^2\v^{-1} - B\v s^2 + A\v^{-1}\G0020 - B\v^{-1}s^2\G2000
\nonumber\\
     &   & - 8Bsc\G1100 - 4B\v(c^2-s^2)\G0200 + 2i\hbar Bsc .
\end{eqnarray}
Then we can obtain the hamiltonian,
\begin{eqnarray}
\underline{\mathscr{H}'}_{\textrm{eff}}
  & = & - p^2 - \G0020 + 3\pi G \Big( \v^2s^2 + s^2\G2000
        + 8\v sc\G1100 + 4\v^2(c^2-s^2)\G0200 - 2i\hbar\v sc \Big)
\nonumber\\
  & = & - p^2 - \G0020 + 3\pi G \Big( \v*(\v s^2) + s^2\G2000
        + 8\v sc\G1100 + 4\v^2(c^2-s^2)\G0200 \Big) .
\end{eqnarray}
So ignoring the higher order terms, there is only one term $\v^2(c^2-s^2)\G0200$ which is different to the one $\v^2\frac{2\hbar^2}{d^2}$ in our method. This difference comes from the mismatching between $\frac{2\hbar^2}{d^2}$ in (\ref{equ:chap4:CohMeth}) which is the fluctuation of $\sin2b$ and $\G0200$ in (\ref{equ:chap4:EffConMeth}) which is the fluctuation of $b$. The reason underlining this differece still needs further investigation.

In conclusion, it shows that the method of effective constraints we used to handle quantum constraint systems, solving unphysical variables and keeping only physical variables to obtain the effective Hamiltonian constraint is reasonable. Additionally, the property that the classical big bang singularity is replaced by the quantum bounce in Loop Quantum Cosmology is in fact reliable.

\begin{acknowledgements}
This work is supported by NSFC (No.10975017) and the Fundamental
Research Funds for the Central Universities.
\end{acknowledgements}

%\section*{Appendix}

%\newpage

\end{document}